\documentclass[manuscript]{acmart}
\renewcommand\footnotetextcopyrightpermission[1]{}
\AtBeginDocument{%
  \providecommand\BibTeX{{%
    \normalfont B\kern-0.5em{\scshape i\kern-0.25em b}\kern-0.8em\TeX}}}

\usepackage{multirow}
\usepackage{multicol}
\usepackage{mdframed}
\usepackage{color, colortbl}
\usepackage{adjustbox}
\usepackage{rotating, graphicx, wrapfig}
\usepackage{subcaption}
\usepackage{supertabular,booktabs}
\usepackage[graphicx]{realboxes}

\definecolor{lightgray}{gray}{0.9}

\mdfdefinestyle{RQFrame}{
 outerlinewidth=0pt,
 skipabove=0pt,
 skipbelow=0pt,
 innertopmargin=6pt,
 innerbottommargin=0pt,
 linewidth=0pt,
 topline=false,
 rightline=false,
 leftline=false,
 innerrightmargin=4pt,
 innerleftmargin=4pt}

\newcounter{RQCounter}
\newcommand{\RQ}[2]{
\refstepcounter{RQCounter} \label{#1}
\begin{mdframed}[style=RQFrame]\noindent
    \textbf{RQ}$_{\arabic{RQCounter}}$.~\emph{#2}
\end{mdframed}
}
\newcommand{\hr}[1]{\textbf{RQ}$_{\ref{#1}}$}
\definecolor{lightgray}{gray}{0.9}

\usepackage{hyperref}

\begin{document}

\title{Integrating Hackathons into an Online Cybersecurity Course}


\author{Abasi-amefon O. Affia}
\affiliation{%
  \institution{University of Tartu}
  \city{Tartu}
  \country{Estonia}}
\email{amefon.affia@ut.ee}

\author{Alexander Nolte}
\affiliation{%
  \institution{University of Tartu}
  \city{Tartu}
  \country{Estonia}}
\affiliation{%
\institution{Carnegie Mellon University}
\city{Pittsburgh}
\country{USA}}
\email{alexander.nolte@ut.ee}

\author{Raimundas Matulevi\v{c}ius}
\affiliation{%
  \institution{University of Tartu}
  \city{Tartu}
  \country{Estonia}
  }
\email{rma@ut.ee}

\begin{abstract}
Cybersecurity educators have widely introduced hackathons to facilitate practical knowledge gaining in cybersecurity education.
Introducing such events into cybersecurity courses can provide valuable learning experiences for students. The nature of the hackathon format encourages a learning-by-doing approach, and the hackathon outcomes can serve as evidence for students knowledge, capability and learning gains.
Prior work on hackathons in education mainly focused on collocated hackathon events in the traditional classroom setting. These hackathon events often took place as a one-off event at the end of the course. 
However, one-off hackathon events at the end of a course might not be sufficient to improve learning. Instead, we focus on analyzing the integration of a series of online hackathon events into an online cybersecurity course and explore how this integration can address online education issues by encouraging collaboration and developing a practical understanding of the delivered course by solving real-world challenges.
We evaluate interventions to foster learning and analyze its effect on collaboration and learning gains for students in the course.
Our findings indicate that students attribute learning benefits to the introduced interventions that supported teamwork and collaboration, maintained student participation and interest in the course, and encouraged learning-by-doing.
\end{abstract}



\maketitle

\section{Introduction}
Cybersecurity education has seen an increased interest in literature and practice because of the alarming rate of cybersecurity breaches due to existing security vulnerabilities~\cite{fowler2016data,key2017automotive}, the reported lack of security experts or shortage of cybersecurity workers to defend against cyber attacks~\cite{boopathi2015learning,weiss2015teaching} and the increase of information and communication technologies in everyday use~\cite{venter2019cyber}. These point to a need to improve existing cybersecurity curricula to foster practical cybersecurity knowledge.
To that end, researchers have proposed expanding the practical aspect of cybersecurity education. One approach that has gained momentum in this regard are educational hackathons organized to support cybersecurity education ~\cite{weiss2015teaching,kharchenko2016university,boopathi2015learning}.
Sadovykh \textit{et al.}~\cite{sadovykh2019hackathons} and Steglich \textit{et al.}~\cite{steglich2020hackathons} report how adding hackathons to an educational curriculum fosters student familiarity with different technologies and supports students in adopting problem-solving practices. 
Hackathons are time-bounded events during which participants with diverse backgrounds form teams and work on projects of interest to them~\cite{pe2019designing}. Hackathons have since begun as competitive coding events and have proliferated into various domains, including corporations~\cite{moe2021improving,pe2020corporate,komssi2015hackathons}, entrepreneurship~\cite{medina2021supporting,richter2021digital,bubbar2019promoting}, civic engagement~\cite{hartmann2018innovation,lodato2016issue,henderson2015getting}, communities~\cite{nolte2020support,pe2019understanding,huppenkothen2018hack} and others.
Educational settings, in particular, have been found to benefit from the introduction of hackathons because they encourage students to practice the concepts learned in the classroom~\cite{gama2018hackathons,porras2019code}. Consequently, educators have adopted hackathons in traditional education settings related to computer science, software engineering, and STEM~\cite{Seidametova2022hack}.

Current works mainly focus on one-off events at the end of the course.
However, the learning process cannot reach its full potential, especially with online education, until students are encouraged to frequently practice what they learn~\cite{dhawan2020online}. Introducing one-off opportunities at the end of the course to practice what they learnt may not be effective.
For students to excel in their respective courses, they need to be cognitively and socially engaged throughout their course activities since learning is a social and cognitive process~\cite{tinto2011taking}.
A way to achieve this is through collaborative problem solving and active learning approaches, integrated throughout the course, to engage students cognitively in the coursework, develop their learning skills, and enhance academic achievement and engagement~\cite{unal2021effect}. Thus, we can carefully design and introduce hackathons at multiple points in the course to provide intended benefits of improved student interest and engagement, increased interaction between instructor and student, and the opportunity to practice or apply theoretical concepts taught. Our proposed hackathon design covers actions employed to address issues with the learning process of an  online format by introducing  interventions.

Additionally, prior works have also focused on in-person hackathon events in the context of traditional courses~\cite{la2017engineering,tandon2021using}. The COVID-19 pandemic has forced education online. And as a result, online instruction and collaboration are gaining popularity to support the learning and innovative process for students~\cite{dhawan2020online}; but challenges exist. Such challenges are associated with current technological constraints (download errors, installation issues, audio and video problems etc.), reduced student interest and engagement in online study, lack of a sense of belonging and connectedness, distractions at home, and time management challenges for students~\cite{dhawan2020online,gama2021online}. Additionally, the online course can be all theoretical, which does not prompt students to practice alongside the instructor, leading to an inadequate two-way interaction between instructor and student that discourages an effective learning process.

In this paper, we thus investigate:
\begin{center}
    \textit{How can different interventions at online educational hackathons influence learning about cybersecurity?} 
\end{center}
Building on prior work on hackathon interventions that foster learning in cybersecurity~\cite{abasi2020developing}, we explore these hackathon interventions in an online-led cybersecurity course.
Specifically, the questions guiding this research are:
\RQ{RQ1}{How do the interventions contribute to teamwork?}
\RQ{RQ2}{How do the interventions contribute to learning?}
We introduced and integrated a series of online educational hackathons  throughout an online-led cybersecurity course to answer these research questions. We designed hackathon interventions to support teamwork and collaboration, maintain student participation and interest in the course, and encourage learning-by-doing within the online context. Our findings indicate that the hackathon interventions contributed to learning gains, fostered teamwork, helped students maintain interest in the course topic and supported participation.

We organized the rest of our paper as follows: First, we explored the research gap and hackathon design aspects for learning in ~\hyperref[Sec:background]{Section~\ref{Sec:background}}, then explained the setting of the hackathon format and the introduced interventions in ~\hyperref[Sec:researchmethod]{Section~\ref{Sec:researchmethod}}. In~\hyperref[Sec:findings]{Section~\ref{Sec:findings}}, we analyze the student perception of the contribution of the intervention to teamwork and collaboration (\hyperref[RQ1]{\hr{RQ1}}) during these events. Lastly, we discuss how the hackathon interventions contribute to student learning (\hyperref[RQ2]{\hr{RQ2}}) in~\hyperref[Sec:discussion]{Section~\ref{Sec:discussion}}.

\section{Background}\label{Sec:background}
In this section, we will discuss hackathons in cybersecurity education in~\hyperref[Sec:hackandcyber]{Section~\ref{Sec:hackandcyber}}, the design of hackathon interventions introduced in our research in~\hyperref[Sec:designaspects]{Section~\ref{Sec:designaspects}} and prior work in educational hackathons in~\hyperref[Sec:relatedworks]{Section~\ref{Sec:relatedworks}}.

\subsection{Hackathons and Cybersecurity Education}\label{Sec:hackandcyber}
Hackathons have been widely introduced to facilitate training and cybersecurity awareness within the computer science and software engineering communities~\cite{kharchenko2016university,boopathi2015learning,weiss2015teaching,foley2018science}.
However, hackathons in cybersecurity typically focus on the investigation of a topic, tool or technology within the cybersecurity domain and the development of research outcomes, security artefacts and prototypes 
~\cite{kharchenko2016university,boopathi2015learning,weiss2015teaching,foley2018science}.

Kharchenko~\textit{et al.}~\cite{kharchenko2016university} presented a series of hackathons to encourage cybersecurity research, development and university-industry cooperation. The first hackathon covered presentations and brainstorming to develop a platform for  testing the
cybersecurity features of industrial programmable logic controllers (PLCs), forming the basis for future hackathons. Subsequent hackathons covered cybersecurity training, idea generation for  securing field-programmable gate array (FPGA)-based PLCs,  startup development for security testing services, and further actions to support research at the university. 
Foley~\textit{et al.}~\cite{foley2018science} discussed the outcome of a hackathon organized for researchers to secure cyber-physical systems (CPS) using transverse use-cases on shared CPS testbed platforms. Researchers were able to explain ongoing research in securing CPS using the testbeds as a case, learn and teach each other about the technology and their research, and then develop prototypes. 
Lastly, Weiss~\textit{et al.}~\cite{weiss2015teaching} reported the design and experience of using an interactive cybersecurity scenarios framework to teach security analysis and support the computer science curriculum. The paper proposed scenarios to be used by educators to nurture the development of security analysis skills in students to complement theoretical security concepts and specific software tools taught within the school's educational curriculum.
Though linked to universities and an educational curriculum, these papers~\cite{foley2018science,kharchenko2016university,weiss2015teaching} were not designed to support a specific cybersecurity course.

But similar to our proposal, Boopathi~\textit{et al.}~\cite{boopathi2015learning} introduced a training course to learn about cybersecurity using a format that introduces a hackathon much like a capture-the-flag (CTF) competition. Boopathi~\textit{et al.} presented the application of a training format to support graduate and undergraduate level curricula. Here learning happens in three rounds representing three aspects of the training format. The learning round introduced cybersecurity concepts. The jeopardy round tested the participants' knowledge by administering questions to solve as assignments. The interactive rounds tested the gained knowledge in a real-world scenario through gamification. Boopathi~\textit{et al.} thus introduced a hackathon as part of the third round -- but only as a one-off event. Instead, we propose the integration of hackathons into an online cybersecurity course, not as a one-off event but as a series of events, introduced at strategic points of the course to provide ample opportunity for students to learn by doing.

\subsection{Hackathon design aspects for learning}\label{Sec:designaspects}
Online-led instruction and course delivery can introduce challenges in student engagement, collaboration, and even learning within the course~\cite{tinto2011taking}. Thus, online educational hackathons should be designed to promote active, experiential learning through dealing with real-world problems while increasing student interest and engagement within the course~\cite{horton2018project}.

Hackathon interventions can serve to promote engagement, innovation, teamwork, and problem-solving
\cite{rennick2018engineering,abasi2020developing}. 
Based on a prior study on developing hackathon interventions to foster learning~\cite{abasi2020developing}, we can outline the following design aspects in our research that inform interventions suitable for our case.
At a typical hackathon event, organizers tend to devote the early part of any hackathon to problem-set development and planning~\cite{stoyanov2007effect}.
Working on a problem set can require single or group participation, with each participant having the necessary skill or experience to solve the posed problem. Thus, groups or teams of two or more participants contribute to enhancing learning through working together to solve the developed problem, complete needed tasks and learn new concepts~\cite{abasi2020developing}.
However, working in teams can pose challenges, such as the clarity of the team goal, effectiveness of the team process, and level of team participation. To alleviate these side effects, educators should introduce \textit{team management} interventions.
Team management interventions improve teamwork and encourage learning by enhancing collaborative team power.

Problem-solving, even in hackathons, requires that students apply known concepts to develop suitable solutions to set problems~\cite{stoyanov2007effect} and the opportunity to acquire new information relevant to carrying out a specific project. Thus, domain-specific knowledge encourages learning by problem-solving using the hackathon format. This intervention in the educational setting is strongly supported as courses readily provide domain knowledge through \textit{lectures} and related lecture materials. 

Lastly, lectures and lecture materials provide passive learning, where students are not required to be actively involved. Thus, introducing an opportunity for \textit{feedback} encourages an environment of shared inquiry between students and educators to foster learning and creativity and enhance active and collaborative learning activities~\cite{phillips2005strategies}. Feedback given by educators acting as mentors support students to seek their solutions, and findings in the context of educational mentoring attribute increased self-confidence to a positive feedback experience~\cite{nolte2020support}.

\subsection{Related Works}\label{Sec:relatedworks}
Studies have presented results or experiences concerning educational hackathons.
For example, Tandon~\textit{et al.}~\cite{tandon2021using} and Mtsweni~\textit{et al.}~\cite{mtsweni2015stimulating} investigated how educational hackathons can increase student interest levels.
Tandon~\textit{et al.} explored how educational hackathons can increase interest in STEM education and showed positive results in growing participant interest and improving participant knowledge levels~\cite{tandon2021using}. The research suggested benefits from providing an un-intimidating environment by introducing hackathon approaches where students learn by doing and give the participants collective creative liberties to determine the solutions to problems not usually present in the classroom. 
However, this study focuses on collocated hackathons in a traditional classroom course and does not account for the case of online education and collaboration. 
Mtsweni~\textit{et al.} also presented findings to show that the hackathon approach can stimulate and maintain students' interest in computer science with distance teaching mode. The key elements of the proposed hackathon approach covered collaborations, networking, mentoring, hands-on engagement projects, and community involvement~\cite{mtsweni2015stimulating}. The authors conducted hackathons in a hybrid format (physically or online); however, they did not introduce hackathons in a course curriculum.

Hackathons should also foster a rapid learning process for educational courses.
La Place~\textit{et al.}~\cite{la2017engineering} showed how hackathons foster rapid learning-by-doing for engineering students. The research work documented the perceived learning process of the participants and the actions of their learning methods to determine how hackathons can help improve project-based learning courses in engineering to bridge the gap between self-directed active learning and formal learning. According to La Place \textit{et al.}~\cite{la2017engineering}, hackathon attributes to improve learning-by-doing include creative freedom and motivation for problem-solving, team collaboration opportunities, proper team management to achieve the team goals and the constraint of time to complete the hackathon tasks. Although the research discusses valuable attributes of the hackathon format that was beneficial to improve rapid learning, it also focuses on collocated hackathons in a traditional classroom course. 

Recently, some research works have explored remote educational hackathons. For example, Steglich~\textit{et al.}~\cite{steglich2021online} explored how intense collaboration takes place between student teams in an educational hackathon to produce a technological solution and develop professional skills in the online context. The study presents findings that the most valuable skills to enable effective student teams were communication, collaboration, initiative, and creativity/innovation~\cite{steglich2021online}. However, this study did not account for learning gains in this educational hackathon but rather professional skills and collaboration.
Gama~\textit{et al.}~\cite{gama2021online} also presented an experience report on how online educational hackathon is used as a resource to engage students in the development of their semester project. The paper describes details of the students' perception of the hackathon approach used and how it helped to create an intense collaborative experience while having the sense of being virtually collocated~\cite{gama2021online}. However, the authors introduced hackathon interventions at the end of the course and did not focus on increasing engagement or learning during the course instruction.
Lastly, Goodman~\textit{et al.}~\cite{goodman2020learn} proposed a framework for online educational hackathons by generalizing two frameworks for hackathons and CTFs, respectively. The approach focused on three main stages - Learn, Apply, and Reinforce/Share where educators teach students or allow students to self-learn new material, apply gained knowledge and skills to tackle a specific problem (self-refined or introduced by educators), and share/present work done (individually or in teams) at the end of the hackathon. However, the work lacked details or concrete usage of this framework in a course setting.

Our work is thus different from prior studies because we integrated multiple online hackathon events into a cybersecurity course instead of a typical one-off hackathon event at the end of the course. 
Additionally, we evaluate the student's perception of the hackathon format, the designed interventions integrated into the course, the progression of team aspects during these hackathon events, and how these contribute to students' learning gains.

\section{Empirical Method}\label{Sec:researchmethod}
To answer our research questions (\hyperref[RQ1]{\hr{RQ1}}, \hyperref[RQ2]{\hr{RQ2}}), we conducted an action research study~\cite{lewin1946action} applying proposed interventions at a series of hackathon events integrated into an educational course to foster learning. We conducted the study over six (6) months, spanning three (3) course-integrated hackathon events.
We describe the proposed interventions to support the hackathon events in~\hyperref[Sec:interventions]{Section~\ref{Sec:interventions}}, the course design and timeline of the hackathon events in~\hyperref[Sec:settings]{Section~\ref{Sec:settings}}, our data collection activities in~\hyperref[Sec:datacollection]{Section~\ref{Sec:datacollection}} and our analysis procedures in~\hyperref[Sec:analysisprocedure]{Section~\ref{Sec:analysisprocedure}}.

\subsection{Proposed Interventions} \label{Sec:interventions}

Based on prior work~\cite{sadovykh2019hackathons,abasi2020developing} we developed and introduced three interventions to stimulate collaborative problem solving and encourage learning during the hackathon events -- \textbf{\textit{lecture}}, \textbf{\textit{feedback}} and \textbf{\textit{team management plan}} interventions.   We based these interventions on the design aspects previously discussed in \hyperref[Sec:designaspects]{section~\ref{Sec:designaspects}}. They are in line with the framework for online educational hackathons proposed by Goodman~\textit{et al.}~\cite{goodman2020learn}.

First, as part of the cybersecurity course, we provided lecture materials suitable for online instruction. This preparation involved the preparation of video lectures that students can watch online at their convenience and the introduction of discussion sessions and lecture live streams to understand the lecture materials before the hackathon events. We introduced the \textbf{\textit{lecture intervention}} to enable students to learn about basic concepts and techniques inspire students to reflect on their selected use-case and how the security concepts introduced may apply to the use-case. The lecture intervention also provides students with a base knowledge to attempt hackathon tasks during the hackathon event.

We organized ample team interaction with course instructors through online feedback sessions during each hackathon event and feedback to write-ups submitted after each hackathon event. We introduced the \textbf{\textit{feedback intervention}} sessions to provide students with the opportunity to gain expert feedback and answers to questions regarding hackathon tasks. Students were also required to submit write-ups at the end of each hackathon, and the course instructors provided feedback on the write-ups.
Since we designed closely related hackathon events, we hoped that the feedback provided to the last hackathon tasks would benefit the upcoming hackathon.

Lastly, as the students formed teams for the hackathon events, we prepared the \textbf{\textit{team management plan intervention}} documents to help teams plan tasks, work together and complete tasks more efficiently. The team management plan aimed to help students document task assignments, assign responsibilities to the hackathon tasks, and specify deadlines for completing the hackathon tasks. The team management plan also aimed to support cooperative working relationships, task organization and assignment, or team leadership.

\begin{figure*}[hbt!]
  \centering
  \includegraphics[width=0.88\linewidth]{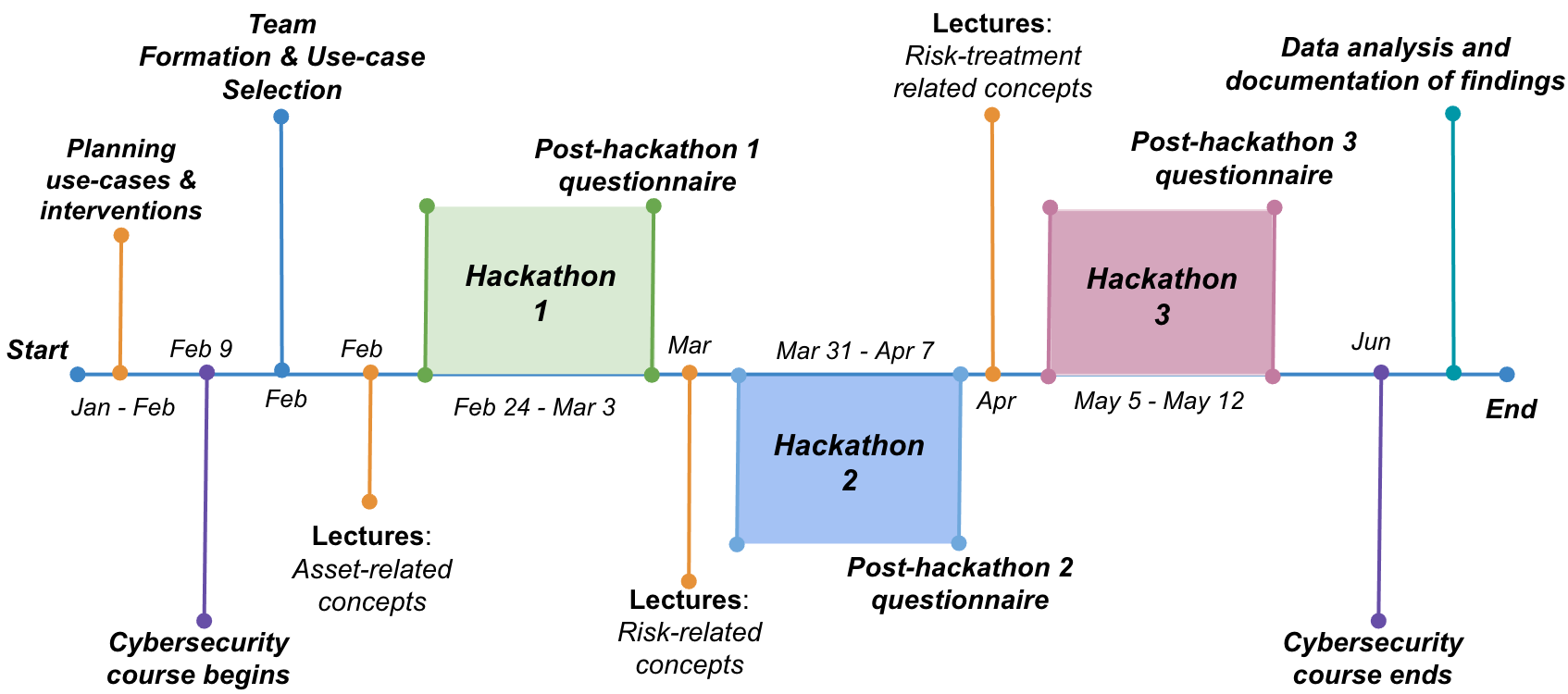}
  \caption{Timeline of activities} \label{Fig:timeline} 
\end{figure*}

\subsection{Setting} \label{Sec:settings}
This section introduces the course design and how we integrated hackathon events into the course. Based on the hackathon planning kit~\cite{nolte2020organize}, we set up the core components of the hackathon events.

\subsubsection{Cybersecurity Course Design}
We introduced our hackathon format as part of a cybersecurity course to teach students about the principles for secure software design from a security risk-aware perspective.
The main goal of the course focused on the security risk management of software systems. The course informed on how to ensure the security of software system assets, security requirements engineering and modelling, and understanding major security controls, like role-based access control and cryptography, fundamental to secure software design. 
As a learning outcome of completing the course, the students should confidently identify causes and consequences of (lack of) system and software security, master techniques to address system and software security problems, elicit and justify the introduction and application of security requirements and controls.

There are three (3) important aspects of ensuring secure software system design concerning security risk management and following the security risk management method used in the course: \textit{(i) asset-related concepts}, \textit{(ii) risk-related concepts}, \textit{(iii) risk-treatment related concepts}. These major aspects led to the context of each of our hackathon events. 
To promote active learning and achieve the course's learning outcome, we proposed hackathon events covering the three (3) aspects of ensuring risk-aware secure software design. This included hackathons focusing on the \textit{asset-related}, \textit{risk-related} and \textit{risk treatment-related} concepts taught  in the lectures. The proposed format enabled us to easily introduce and integrate hackathon events to allow students to understand the theoretical knowledge and apply it in real-world cases.

\subsubsection{Timeline of events}
This section covers the timeline of major events throughout the hackathon integrated course duration also outlined in~\hyperref[Fig:timeline]{Fig.~\ref{Fig:timeline}}.

Before starting the course, we prepared two use-cases for the students to run a security risk analysis on at the hackathon using knowledge gained during the course. 
The use-cases covered two (2) software-intensive systems -- a Bike Sharing System and an Autonomous Vehicle Parking System -- reflecting real-world scenarios to guide the students' learning process. The Bike Sharing System provides bike-sharing services to its users through its major components: Smart Bike (SB), Bike Share Website (BWA), and the Bike Mobile Application (BMA). The Autonomous Vehicle Parking System provides parking services to its users through its major components: Autonomous Vehicle (AV), Parking Service Provider (PSP), and Parking Lot Terminal (PLT). We prepared use-case information, including UML diagrams to show system interaction between components and textual descriptions to explain the system functions 
as much information as possible. We also validated the use-case information gathered through the system stakeholders to confirm that they reflect the real-world software system (to a reasonable degree).
We also prepared the proposed interventions as discussed in~\hyperref[Sec:interventions]{section~\ref{Sec:interventions}}.

Once the course began (see~\hyperref[Fig:timeline]{Fig.~\ref{Fig:timeline}}), we formally introduced the course and its learning outcomes, the course design and the hackathon format to the students. We requested the students to form teams of three (3) or four (4) members and select a preferred use-case for the team. The students freely formed teams and selected their preferred use-case to be analyzed during the hackathon events. Additionally, we provided the first round of lectures introducing the analysis of asset-related concepts of software systems.

We organized the first hackathon event (\textit{Hackathon 1} in ~\hyperref[Fig:timeline]{Fig.~\ref{Fig:timeline}}) about two (2) weeks after the start of the course. We provided the hackathon tasks for students to analyze the~\textit{asset-related concepts} of the chosen use-case with knowledge provided by the lecture intervention -- lectures and lecture resources offered. The students completed the tasks in teams and submitted a security asset analysis document as their hackathon task report. The students also participated in presentation sessions to discuss the outcome of their hackathon tasks. We provided feedback to the students concerning the presented work and the submitted report after the hackathon event and offered feedback on an ad-hoc basis based on requests by the students or teams.

We organized the second hackathon event (\textit{Hackathon 2} in ~\hyperref[Fig:timeline]{Fig.~\ref{Fig:timeline}}) about a month after the first hackathon event. Before the second hackathon, we provided the second round of lectures analyzing \textit{risk-related concepts} of software systems before the event. Thus, we provided hackathon tasks to the students to analyze the~\textit{risk-related concepts} of their chosen use-case building on their asset analysis from the first hackathon event. We introduced the team management plan intervention at this event to support teamwork and collaboration and to help the teams achieve their hackathon tasks more efficiently. The students also participated in online feedback (intervention) sessions with the educators to discuss progress or challenges with their hackathon tasks. At the end of the hackathon event, we asked the teams to submit their team management plan document alongside their task report, documenting their completed ~\textit{risk-related} analysis, as an output of the hackathon event. We provided additional feedback during presentation sessions and written feedback to the task report submitted after the hackathon event. 

We organized the third and final hackathon event (\textit{Hackathon 3} in ~\hyperref[Fig:timeline]{Fig.~\ref{Fig:timeline}}) about a month after the second hackathon event. Before this event, we provided the third round of lectures analyzing \textit{risk treatment-related concepts} of software systems before the event, where we emphasized the modelling of role-based access controls. Thus, the hackathon tasks for the third hackathon event covered an analysis of the~\textit{risk treatment-related concepts} of their chosen use-case building on the risk analysis from the second hackathon event. We provided multiple online feedback sessions to discuss progress or challenges with the hackathon tasks. The students completed and submitted a cumulative task report of the hackathon tasks from all three (3) hackathon events, 
representing a security analysis of the use-case they selected (Bike Sharing System or an Autonomous Vehicle Parking System). As an output of the hackathon tasks, the teams provided in the report a proposal for secure software design following \textit{asset-related}, \textit{risk-related}, and \textit{risk treatment-related} analysis of the selected use-case. This report counted to the student's grade at the end of the course.

\subsection{Data collection}\label{Sec:datacollection}
After each hackathon event, we conducted a post-hackathon questionnaire using pre-existing instruments that we adapted for our study.~\hyperref[tab:Qinstruments]{Table~\ref{tab:Qinstruments}} in ~\hyperref[sec:app]{Appendix~\ref{sec:app}} shows the scales we utilized for our questionnaire instruments. We collected student perception of team familiarity once at the first hackathon event because continual data collection was unnecessary for this data point as each member remained in the same team for all hackathon events. We also collected student perception of learning only at the final hackathon event, allowing the student to respond retrospectively to the hackathon events and the course.
The questionnaire also covered the students' perception of usefulness and satisfaction with the three interventions and their contribution to team properties such as team familiarity, goal clarity, team efficiency, and team collaboration, thus answering~\hyperref[RQ1]{\hr{RQ1}}. We also collected data on the students' perception of learning achieved to answer~\hyperref[RQ2]{\hr{RQ2}}.
In addition, we asked open-ended questions in the questionnaire to provide more contextual information about the students' perception of the team experience and evaluate how the different interventions benefited the teamwork and the learning process (\hyperref[RQ1]{\hr{RQ1}}, \hyperref[RQ2]{\hr{RQ2}}).

\subsection{Analysis procedure}\label{Sec:analysisprocedure}
After the cybersecurity course completion, we analyzed data collected from each hackathon event questionnaire and qualitatively analyzed open-ended questions to support arguments and provide potential explanations to our analysis from the questionnaire scales and answer research questions~\hyperref[RQ1]{\hr{RQ1}} and \hyperref[RQ2]{\hr{RQ2}}.
We selected responses from six (6) teams for our data analysis based on the team size (between three (3) and four (4) members), course grade outcome and teams who provided more complete responses to the questionnaires. We selected and grouped the highest-scoring, middle and lower scoring teams in the course grade outcome criteria. 
Our selections resulted in three (3) major groups by grade selection having two (2) teams per grade group. The selected team characteristics are summarized in~\hyperref[tab:teamcharacteristics]{Table~\ref{tab:teamcharacteristics}}.

\begin{table}[H]
\caption{Team characteristics}
    \label{tab:teamcharacteristics}
    \centering
\begin{tabular}{|p{0.23\linewidth}|p{0.15\linewidth}|p{0.3\linewidth}|}
\hline
Grade selection      & Teams  & Participants \\ \hline
\multirow{2}{*}{High grade}   & Team A & A01, A02, A03, A04    \\ \cline{2-3} 
                              & Team B & B01, B02, B03, B04    \\ \hline
\multirow{2}{*}{Middle grade} & Team C & C01, C02, C03, C04    \\ \cline{2-3} 
                              & Team D & D01, D02, D03, D04    \\ \hline
\multirow{2}{*}{Low grade}    & Team E & E01, E02, E03, E04    \\ \cline{2-3} 
                              & Team F & F01, F02, F03    \\ \hline

\end{tabular}
\end{table}

\subsubsection{Pre-processing}
We first prepared the data collected, replacing the 5-point scale answers with corresponding numbers from $1-5$, accounted for negative scales and reversed them accordingly. The questions were also coded for easy reference and use in further preparation stages and analysis. 
\begin{figure*}[hbt!]
  \centering
  \includegraphics[width=1\linewidth]{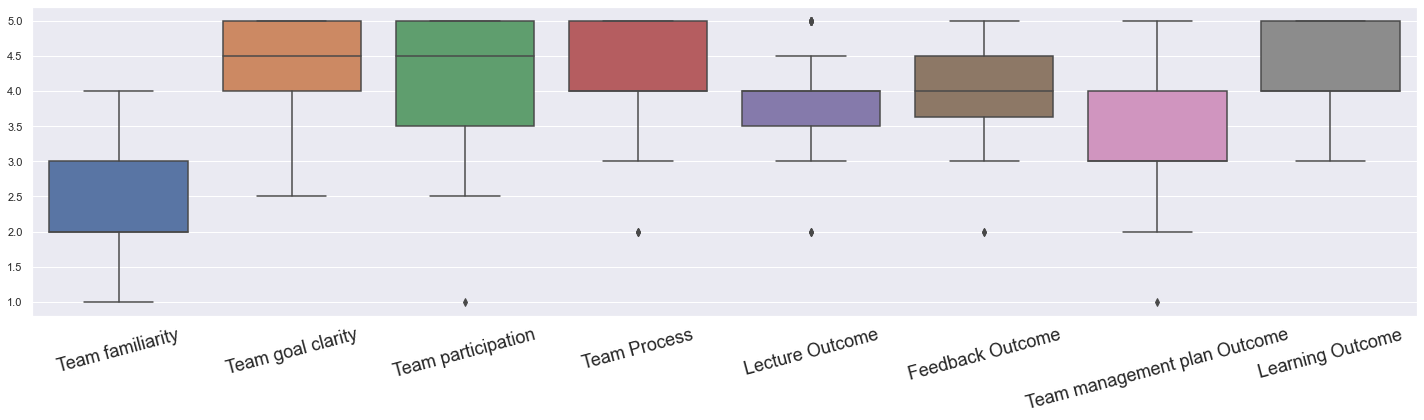}
  \caption{Data points collected for analysis from selected teams.} \label{Fig:datapoints} 
\end{figure*}
\subsubsection{Reliability}
We tested the reliability of the variables using Cronbach's Alpha.
We created summated scales by calculating a composite score from each Likert-type item and gave it a human-readable name format. For example, naming summated scale questions that measure how well the student participants knew each other before the hackathon activities as the \textit{Team familiarity} scale.
We will continue to use these summated scales and not individual question items for further analysis of the data as Cronbach's alpha does not provide reliability estimates for single items~\cite{gliem2003calculating}. We presented results for the reliability of the items in \hyperref[tab:itemvalidity]{Table~\ref{tab:itemvalidity}} for the selected sample. The Cronbach's Alphas were higher than the recommended value of $0.70$, which corroborates reliability.

\subsubsection{Descriptive analysis}
Once we found our scales reliable, we then grouped individual responses to those scales into their respective teams and grouped each team into their grade groups for additional analysis.
Descriptive statistics used for our data items include the median for central tendency and interquartile range as a measure of statistical dispersion.
These methods represented the data collected for all teams and aided the extraction of valuable observations. We illustrate the median and interquartile range values per team (see \hyperref[tab:allprop]{Table~\ref{tab:allprop}}).

\begin{table}[H]
\caption{Scale reliability.}
    \label{tab:itemvalidity}
    \centering
\begin{tabular}{p{0.4\linewidth}p{0.25\linewidth}}
\hline
     & Cronbach's Alpha ($\alpha$)  \\ \hline
Team familiarity &  0.77 \\ \hline
Team goal clarity & 0.76 \\ \hline
Team participation & 0.88 \\ \hline
Team process & 0.91 \\ \hline
Feedback outcome & 0.83 \\ \hline
Lecture outcome & 0.95 \\ \hline
Team management plan outcome & 0.91 \\ \hline
Learning outcome & 0.89 \\ \hline
\end{tabular}
\end{table}

\begin{table*}[h]
\centering
    \caption{Calculated cumulative data points from all three (3) hackathon events (response count $n$, median $M$, and interquartile range $IQR$) used in qualitative analysis. Median and interquartile range values are from responses given on a 5-point scale. \\
    * We collected team familiarity data once at the first hackathon events. Continual data collection was unnecessary for this data point as each member remained in the same team for all hackathon events.\\
    **Learning data was collected once at the last hackathon event.}
    \label{tab:allprop}
     \resizebox{0.95\columnwidth}{!}{%
\begin{tabular}{|p{0.065\linewidth}|p{0.06\linewidth}|p{0.03\linewidth}|p{0.09\linewidth}|p{0.07\linewidth}|p{0.08\linewidth}|p{0.065\linewidth}|p{0.065\linewidth}|p{0.065\linewidth}|p{0.086\linewidth}|p{0.09\linewidth}|} \hline
Grade \newline selection & \multicolumn{2}{c|}{Descriptive statistic} & Team \newline familiarity\textbf{*} & Team goal clarity & Team participation & Team \newline process & Lecture & Feedback & Team management plan & Learning\textbf{**} \\ \hline

\multirow{6}{*}{High} & \multirow{3}{*}{Team A} & n & 4 & 12 & 12 & 12 & 12 & 12 & 8 & 4 \\ \cline{3-11}
 & & M  & 3.50 & 5.00 & 4.25 & 4.50 & 4.00 & 4.25 & 3.75 & 4.00 \\ \cline{3-11}
 & & IQR & 0.63 & 0.50 & 0.56 & 0.50 & 0.56 & 0.63 & 0.50 & 0.13 \\ \cline{2-11}
 & \multirow{3}{*}{Team B} & n & 4 & 11 & 11 & 11 & 11 & 11 & 7 & 3 \\ \cline{3-11}
 & & M  & 3.00 & 4.00 & 4.00 & 4.00 & 4.00 & 4.00 & 4.00 & 5.00 \\ \cline{3-11}
 & & IQR & 0.13 & 0.75 & 0.75 & 0.75 & 0.25 & 0.00 & 0.25 & 0.25\\ \hline    
 
\multirow{6}{*}{Middle} & \multirow{3}{*}{Team C} & n & 2 & 9 & 9 & 9 & 9 & 9 & 7 & 4 \\ \cline{3-11}
 & & M & 2.00 & 4.50 & 4.00 & 4.00 & 3.00 & 4.00 & 3.00 & 4.00 \\ \cline{3-11}
 & & IQR & 0.00 & 0.50 & 0.25 & 0.00 & 0.50 & 0.50 & 0.25 & 0.25 \\ \cline{2-11}
 & \multirow{3}{*}{Team D} & n & 2 & 8 & 8 & 8 & 8 & 8 & 5 & 3 \\ \cline{3-11}
 & & M & 2.00 & 4.25 & 4.25 & 4.00 & 4.00 & 4.00 & 3.00 & 4.00 \\ \cline{3-11}
 & & IQR & 0.00 & 0.63 & 0.75 & 0.75 & 0.31 & 0.00 & 0.50 & 0.00 \\ \hline
 
 \multirow{6}{*}{Low} & \multirow{3}{*}{Team E} & n & 4 & 11 & 11 & 11 & 11 & 9 & 7 & 3 \\ \cline{3-11}
 & & M  & 2.00 & 5.00 & 5.00 & 5.00 & 4.00 & 4.00 & 3.00 & 4.00 \\ \cline{3-11}
 & & IQR & 0.25 & 0.00 & 0.00 & 0.00 & 0.75 & 0.50 & 0.75 & 0.25 \\ \cline{2-11}
 & \multirow{3}{*}{Team F} & n & 2 & 6 & 6 & 6 & 6 & 5 & 2 & 2 \\ \cline{3-11}
 & & M & 1.00 & 4.00 & 4.50 & 4.00 & 4.00 & 4.00 & 4.50 & 4.50 \\ \cline{3-11}
 & & IQR & 0.00 & 0.19 & 0.69 & 0.19 & 0.38 & 1.00 & 0.25 & 0.25 \\ \hline

\end{tabular}}
\end{table*}

\section{Findings}\label{Sec:findings}
This section outlines the students' perception of each learning intervention for each team and the differences between teams by their team properties and learning process. 
The participants who formed part of the study were cybersecurity students with different levels of prior knowledge related to cybersecurity coming from different backgrounds. We did not examine their background in this paper though. The majority of participants who reported their age (n=23) were less than fifty (50) years, with an average of twenty-eight (28) years.

In general, we observed a positive perception of the team properties, interventions introduced and learning outcome at the end of the course.  From our observations, the medians of student responses to most data points remained above the average of our 5-point Likert scale interval (see ~\hyperref[Fig:datapoints]{Fig~\ref{Fig:datapoints}} and ~\hyperref[tab:allprop]{Table~\ref{tab:allprop}}). However, this was different for team familiarity suggesting lower familiarity between team members. 
Comparing box plots in~\hyperref[Fig:datapoints]{Fig~\ref{Fig:datapoints}}, the data suggests that overall, the students had a high level of agreement on how well the team properties and the interventions benefited them and on learning gains. 
In this section, we analyze collected data to understand better the student's perception of the team properties and the contribution of the interventions to the team properties that foster teamwork in~\hyperref[Sec:teamproppercep]{Section~\ref{Sec:teamproppercep}} to answer \hyperref[RQ1]{\hr{RQ1}}. We also discuss the interventions contributions to learning in~\hyperref[Sec:learnpercep]{Section~\ref{Sec:learnpercep}} to answer \hyperref[RQ2]{\hr{RQ2}}.

\subsection{Perception of Team Properties}\label{Sec:teamproppercep}
\begin{figure}[H]
  \centering
  \includegraphics[width=0.8\linewidth]{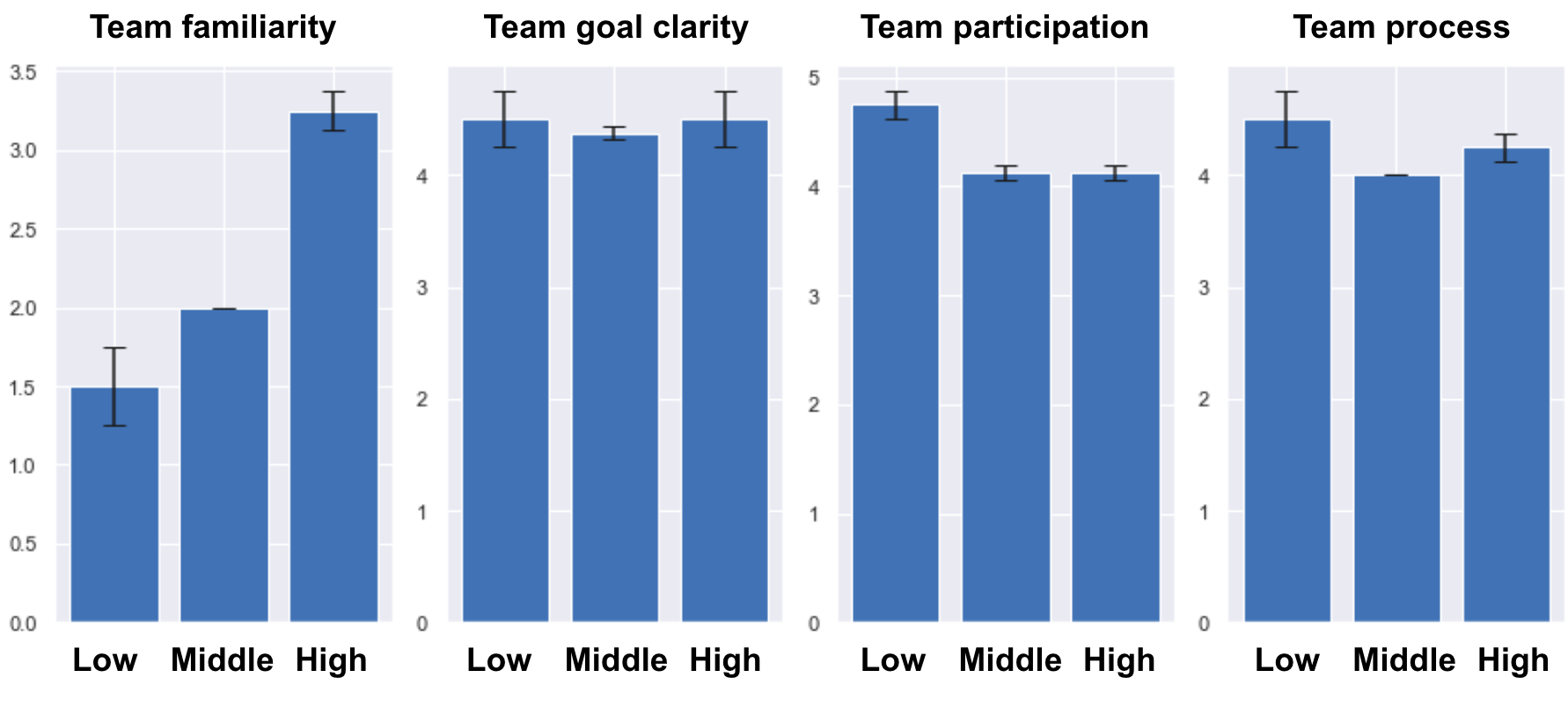}
   \vspace{-5pt}
  \caption{Team properties per grade group.} \label{Fig:teamprop} 
 \vspace{-8pt}
\end{figure}

By analyzing the responses, we observed that the teams formed for the hackathon events were mainly comprised of individuals who were not very familiar with each other as the teams reported lower values of team familiarity (as seen in~\hyperref[Fig:datapoints]{Fig~\ref{Fig:datapoints}}). This observation can be due to various issues, including the diversity of the students taking the course or having trouble meeting before the course or before team formation due to the COVID-19 pandemic. 
However, high grade teams ($M=3.25$, $IQR=0.13$, see \textit{Team Familiarity} in~\hyperref[Fig:teamprop]{Fig~\ref{Fig:teamprop}}) 
possessed an above-average and relatively higher perception of team familiarity than teams in other grade groups, indicating that familiarity among team members was beneficial in having effective teamwork and collaboration for the hackathon tasks (see~\hyperref[Fig:teamprop]{Fig~\ref{Fig:teamprop}} and \hyperref[tab:allprop]{Table~\ref{tab:allprop}}). Student B01, from a high grade team -- Team B, reported to \textit{``already know teammates from the previous semester''}(B01). 

\begin{figure}[H]
  \centering
  \includegraphics[width=0.8\linewidth]{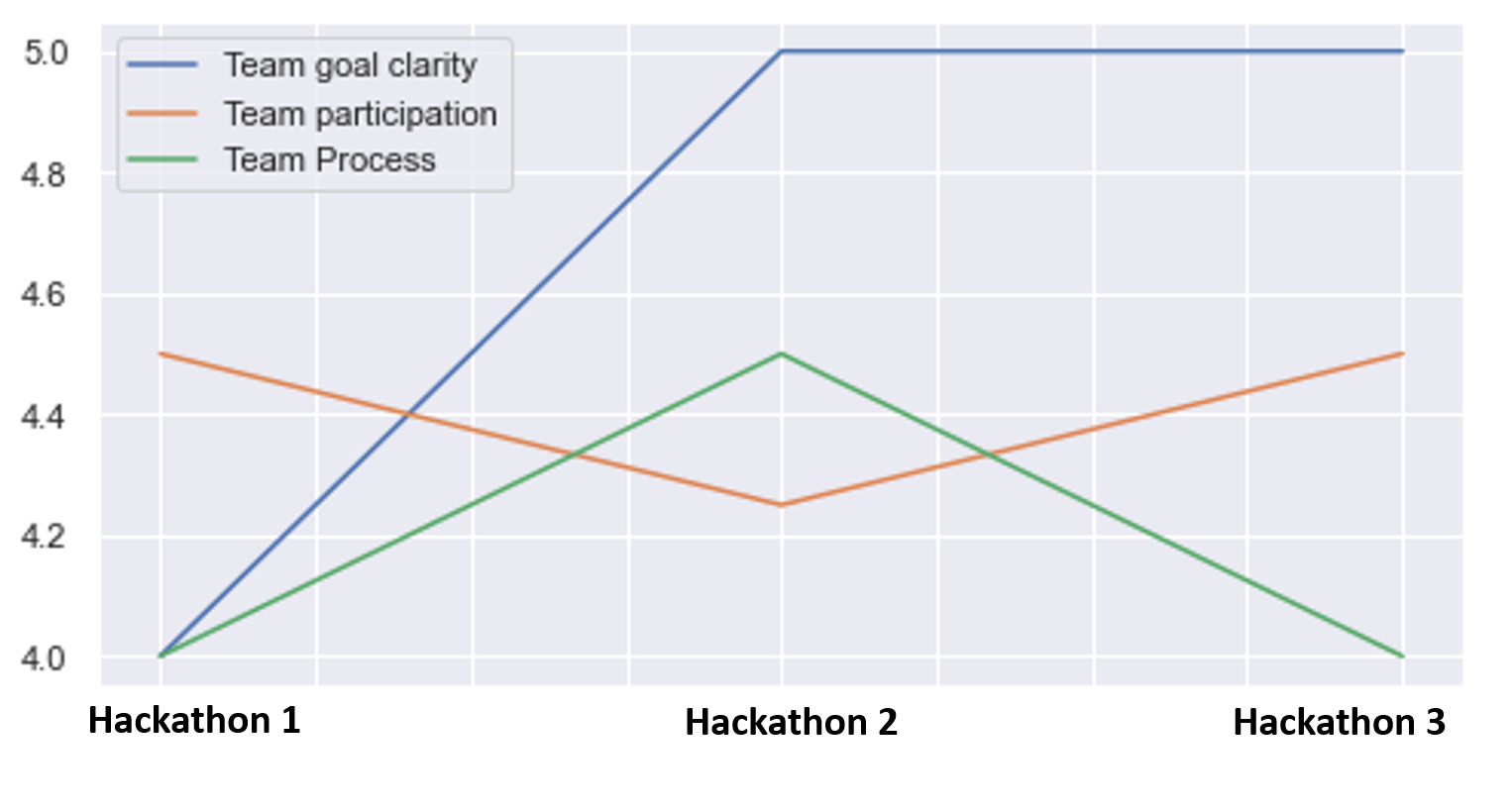}
  \caption{Team properties at each hackathon event.} \label{Fig:teamprophack} 
\end{figure}

All teams reported a higher than average perception of team process with $values \geqq 4.00$.
We observed a 
rise in the team process between the first and second hackathon event and a drop at the third hackathon event (see ~\hyperref[Fig:teamprophack]{Fig~\ref{Fig:teamprophack}}).
The rise in the perceived team process effectiveness may be due to several aspects, including the team members getting to know each other and learning from experiences of the first hackathon event to improve coordination and effects from introduced interventions. However, the drop in perceived team process effectiveness in completing hackathon tasks as seen in ~\hyperref[Fig:teamprophack]{Fig~\ref{Fig:teamprophack}}, may be due to the natural progression of the cybersecurity course and hackathon task difficulty alongside other classes the student will take in parallel during the semester as expected. Student C02, from a middle grade team -- Team C reported that by the third hackathon event it was \textit{``very hard to coordinate team efforts when each and every course requires team work and has different members; ... context switching in human brain does not allow such load''} (C02).
However, this drop in perceived team process effectiveness remained $\geqq 4.00$.

All teams reported a higher than average perception of team goal clarity with $values \geqq 4.00$.
The teams also reported a relatively higher perception of team goal clarity compared to other team measures (i.e., team process, team participation). We observed a significant rise in the team goal clarity between the first and second hackathon event and relative stability by the third hackathon event (see ~\hyperref[Fig:teamprophack]{Fig~\ref{Fig:teamprophack}}).
The rise and relative stability of the perceived team goal clarity values may be due to the team learning from experiences of previous hackathon events to redefine their goals and gain more from the introduced interventions such as the team management plan to help with coordinating duties and responsibilities.

Lastly, all teams reported a higher than average perception of team participation with $values \geqq 4.00$.
We however saw a drop in team participation between the first and second hackathon event, although, with the drop, team participation perception remained $\geqq 4.00$. But, the third hackathon event saw a rise at the third hackathon event to the same level as the first hackathon (see ~\hyperref[Fig:teamprophack]{Fig~\ref{Fig:teamprophack}}). This may be due to the need for increased participation within teams at the first hackathon event to initially define the team goals and tasks, and eventually, increased participation to compile the final security analysis report at the third hackathon event. It is possible that not as much participation was necessary at the second hackathon event.

\subsection{Perception of Learning}\label{Sec:learnpercep}
\begin{figure}[hbt!]
  \centering
  \includegraphics[width=0.8\linewidth]{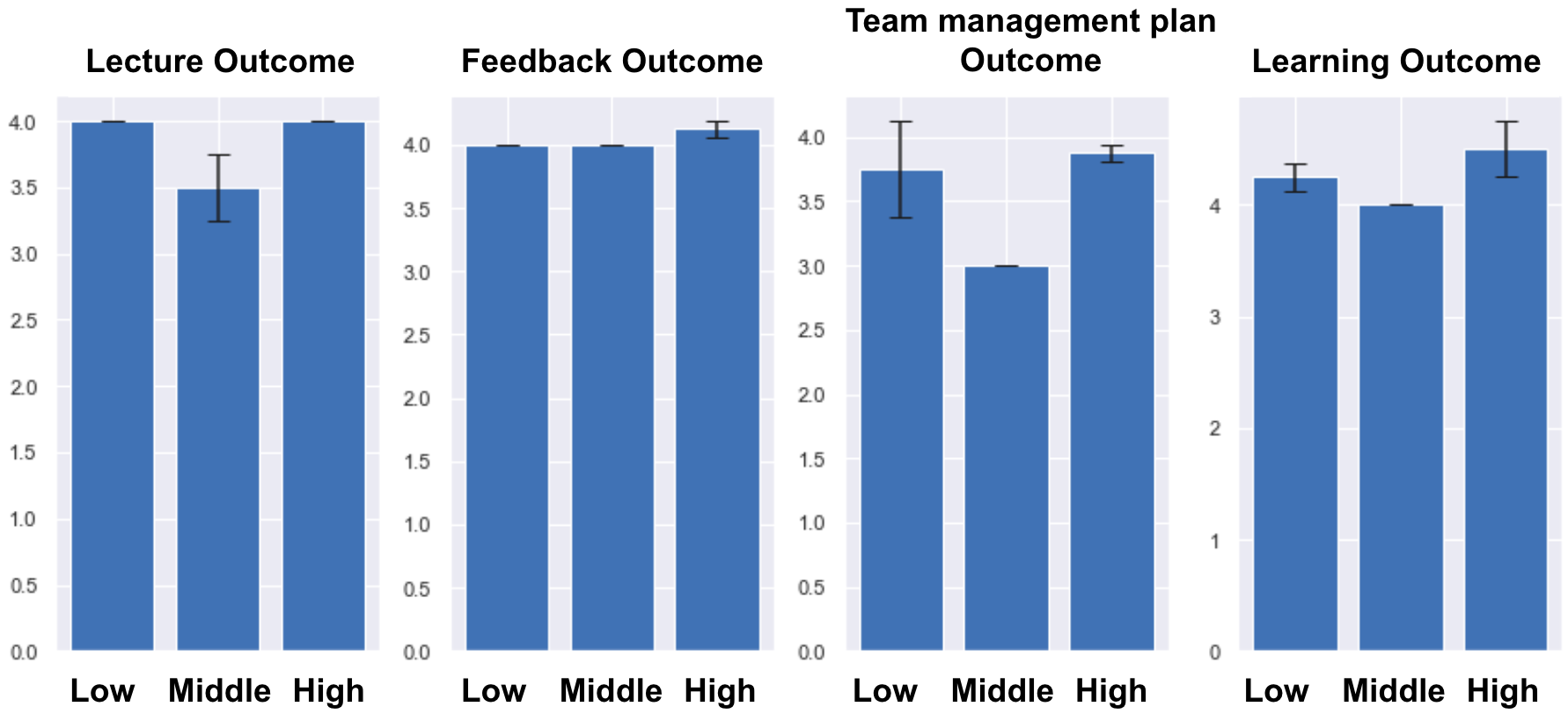}
  \caption{Intervention and learning measures per grade group.}
 \label{Fig:intervenprop} 
\end{figure}

Students reported positive learning experiences in responses to the open-ended questions and analyzed teams reported a higher than average perception of learning. High grade teams reported higher values of achieving the course learning outcome than other grade groups ($M$ = $4.50,IQR=0.25$, see \textit{Learning Outcome} in~\hyperref[Fig:intervenprop]{Fig~\ref{Fig:intervenprop}}), with low grade teams having a reported value of $M=4.25, IQR=0.13$ and middle grade teams having a reported value of $M=4.00, IQR=0.00$ (see \textit{Learning Outcome} in~\hyperref[Fig:intervenprop]{Fig~\ref{Fig:intervenprop}}).
Student B01 added that the hackathon tasks contributed to learning as it \textit{``allowed me to understand some concepts in-depth or concepts that seemed clear in class but were not in reality''}  (B01). 
We also observed that the interventions contributed to perceived learning. Student A02 commented about the "team management plan intervention, stating that the \textit{``team management plan was not easy, however, I believe it provides a positive impact in enhancing my knowledge''}  (A02). Student B01 also added that the \textit{``the lectures contributed to learning''}  (B01), and "textit{``the lecture resources provided additional context to class lectures''}  (B01). 
However, student F01 complained that the lectures had \textit{``too much content and it gets confusing at times''}  (F01).
Regarding the feedback intervention, student B01 noted that the online feedback given helped most for learning as it allowed the team to \textit{``ask for explanations''}  (B01), and "\textit{``better understand some concepts that seemed clear but were not''} (B01).

\subsection{Perception of Interventions} 
We analyzed the students' perception of each intervention -- \textit{lectures}, \textit{feedback}, \textit{team management plan} based on the student grade groups (see ~\hyperref[Fig:intervenprop]{Fig~\ref{Fig:intervenprop}}) to answer \hyperref[RQ2]{\hr{RQ2}}.

\subsubsection{Lecture Intervention}
All teams reported a higher than average perception of the usefulness of the lecture intervention. High grade teams ($M=4.00, IQR=0.00$, see \textit{Lecture Outcome} in~\hyperref[Fig:intervenprop]{Fig~\ref{Fig:intervenprop}}) however, reported higher values of the lecture intervention than other teams followed by low grade teams ($M=4.00, IQR=0.00$) then middle grade teams ($M=3.50, IQR=0.25$).
At the first hackathon event, Student A01 commented on how Team A worked with the lecture intervention. Student A01 stated that although it was personally easy to follow the lectures and lecture resources, \textit{``the most difficult part is to work with the team, as everyone has a different understanding of the lecture''}  (A01). As such, Student A01 asserted that the team had issues with \textit{``correctly completing parts of the hackathon tasks due to these misunderstandings of the lecture''}  (A01). Student F01 indicated about the lecture content that \textit{``sometimes it feels that there is too much content and it gets confusing at times''}  (F01) and Student C02 corroborated this, stating that \textit{``some of the lecture topics needs to be reworked (due to the difficulty of the topics)''}  (C02).
At the second hackathon event, student E01 commented that the \textit{``lectures have touched nicely the aspects that are expected by the team to work on in the hackathon''}  (E01) but also noted that \textit{``...in some cases, it seems we had misunderstood lecture materials''}  (E01). 
Student B01, however, added that for the second hackathon event, \textit{``without the lectures and reading resources it was quite unfeasible completing the hackathon tasks''}  (B01). 
At the third hackathon event, and in retrospect of the hackathon events completed, Student B01 repeated that \textit{``the lectures and reading resources were useful in understanding the hackathon tasks''}  (B01), thus encouraging task completion within the team and an improved team process. 

\subsubsection{Feedback Intervention}
High grade teams ($M = 4.13, IQR=0.06$, see \textit{Feedback Outcome} in~\hyperref[Fig:intervenprop]{Fig~\ref{Fig:intervenprop}}) reported higher values of the feedback intervention than other teams followed by middle grade teams ($M=4.00, IQR=0.00$) then low grade teams ($M=4.00, IQR=0.00$).
For written feedback, Student A03 noted that \textit{``it felt vague and rushed''}  (A03), further clarifying that members of team A \textit{``needed to ask for clarification on every single point individually as to what the feedback comment meant''}  (A03). Student D02 also noted that \textit{``receiving feedback was good, and it helped, but there might be a more effective way of doing it in the future for both students and lecturers''}  (D02). However, student B01 highlighted how the written feedback intervention was helpful for team B, stating that it \textit{``allowed us to find and "correct some inconsistencies with the previous work''}  (B01) and \textit{``correct many errors before the submissions''}  (B01).
Student B01, however, preferred online feedback sessions, stating that \textit{``the opportunity to get explanations in real-time about the feedback was better''}  (B01). 
Student B01 also clarified that for Team B, the online feedback was more useful \textit{``because having a review of the current work at sessions instead of general comments allowed us to find some errors''}  (B01).
Student A02 corroborated this, stating that Team A \textit{``used the online feedback sessions to double-check that we got the tasks right''}  (A02) and student C01 also remarked on how the \textit{``feedback sessions and feedback helped a lot''}  (C01) for the team.

\subsubsection{Team management plan Intervention}
All teams reported a higher than average perception of the team management plan. High grade teams ($M$ = $3.88, IQR=0.06$, see \textit{Team management plan Outcome} in~\hyperref[Fig:intervenprop]{Fig~\ref{Fig:intervenprop}}) reported a higher values related to the team management plan intervention than other teams followed by low grade teams ($M=3.75, IQR=0.38$) followed by middle grade teams ($M=3.00, IQR=0.00$).
%
At the first hackathon event,  we observed a low perception of team familiarity and other issues with team aspects. Student A03 commented on problems with the team experience and organization, noting that \textit{``it was difficult to plan and find the time when all of us were free to work together''}  (A03).
Student B01 (Team B) reported that although the team \textit{``worked quite well''}  (B01), they \textit{``were disorganized''}  (B01) for the first hackathon as they attempted to \textit{``divide the tasks in a fair way at the beginning''}  (B01). We introduced the team management plan intervention at the second hackathon event. Student E01 noted that \textit{``communication and organization of the team was very effective''}  (E01). Student B01 added that the introduction of the team management plan \textit{``was useful to keep track of who is doing what and to see the progress''}  (B01).
But, Team C and Team D (from the middle grade group) reported the lowest values for the team management plan intervention. Student D01 noted that although the team management plan was \textit{``beneficial for clearly defining smaller tasks..., the added value from the management plan was minimal. However, it didn't hinder us or add to our workload''}  (D01).

\section{Discussion}\label{Sec:discussion}
In this section, we will discuss the student's perception of how the interventions benefited teamwork (\hyperref[RQ1]{\hr{RQ1}}) and contributed to student learning (\hyperref[RQ2]{\hr{RQ2}}).

\subsection{Teamwork and Collaboration}
The introduced interventions showed effects on teamwork and collaboration on the hackathon tasks. 
We found that the lecture intervention contributed to the perception of students' effectiveness in completing the hackathon task within the team, especially when the lectures and lecture resources are understandable and applicable to the use-case and the hackathon tasks. The lectures also provided fundamental conceptual security knowledge that was crucial to achieving the hackathon tasks.
The introduction of the team management intervention could have improved the team process and effectiveness to work on the hackathon tasks. We noticed that with the observed perception of team familiarity (see \hyperref[Sec:teamproppercep]{section~\ref{Sec:teamproppercep}}), coordinating tasks within the team could be challenging and sometimes time-wasting as each team member is just learning to collaborate for the first time. The team management plan provided sections that aid definition and coordination of hackathon tasks, assignment of responsibility, and setting deadlines for hackathon tasks where Team B (from the high grade group) and Team F (from the low grade group) benefited the most. Although responses from Team C and Team D (from the middle grade group) can indicate that the team management plan provided minimal contribution (see \textit{Team management plan Outcome} in~\hyperref[Fig:intervenprop]{Fig~\ref{Fig:intervenprop}}), it did not hinder the students from their hackathon tasks nor place an additional workload on the teams. 
Lastly, we saw how the feedback interventions contributed to the teamwork through the opportunity to quickly clear out misunderstandings that may exist within the team about the lectures and the previous and current hackathon tasks. The feedback intervention improves the team effectiveness in completing hackathon tasks correctly. We also saw the contributions of the online feedback sessions over written feedback. Here, an asynchronous form of feedback seemed to improve the teams' ability to handle hackathon tasks quicker without task roadblocks that can be solved by expert guidance in a shorter period.

Our findings thus address the gap identified by Gama\textit{ et al.}~\cite{gama2021online} by designing a hackathon approach that shows results in stimulating student engagement throughout the online course through the hackathon interventions and its introduction at strategic points during the course.

\subsection{Learning Outcome} 
The students reported a positive learning experience in this course. Additionally, we observed that the hackathon format integrated into the course improved learning, allowing the students to practice concepts introduced during the course and get feedback that helps improve knowledge in the domain and their process. 

The lecture intervention was able to provide in-depth information on the security concepts needed for the course. Students reported learning by applying security concepts and practices explained in the lectures in their hackathon tasks. Additionally, the hackathon tasks were crafted with the course curriculum and lecture paths in mind, increasing applicability and the opportunity to learn by doing. The feedback intervention showed greater usefulness in learning as it allowed teams to discuss possible misunderstandings and errors found in the past hackathon task reports and make corrections for future hackathon tasks. The online feedback sessions especially allowed students to discuss current hackathon tasks to prevent repeating past mistakes or introducing new errors into the task outcomes.
The team management plan intervention also served to improve collaborative power for teams, which improves teamwork and encourages rapid learning by working together to complete hackathon tasks correctly.

Although all teams reported positive benefits of the interventions to learn, we can observe how the interventions contributed to learning by analyzing the learning experiences of teams in the high grade group instead of teams in the middle grade or low grade group. We saw that teams in the high grade group reported higher benefits from the interventions than other teams. Responses from members of Team A and Team B (from the high grade group) showed that they could take advantage of the provided interventions. They could do this by applying the lecture resources to understand and complete the hackathon tasks, using the feedback sessions to discuss the written feedback points and other task-related questions, and asking questions concerning past hackathon tasks completed. The teams also used the team management plan to work more effectively, thereby providing a positive impact in enhancing knowledge. The positive benefits are also evident in the teams in the high grade group reported higher learning outcomes than students in low and middle grade groups.

Our findings address the gaps in Steglich~\textit{et al.}~\cite{steglich2021online} by introducing interventions that support collaboration between students in teams alongside interventions to support learning gains. Through our approach, the hackathon format supported learning gains at multiple points in the course, not just at the end of the course, as done in Gama~\textit{et al.}~\cite{gama2021online}. Additionally, we address the gaps in La Place~\textit{et al.} \cite{la2017engineering} and Tandon~\textit{et al.}~\cite{tandon2021using} to demonstrate how the introduced interventions support learning-by-doing through educational hackathons in an online context.

\subsection{Suggested Improvements}

Our findings on the lecture intervention indicate that an understanding (or lack of understanding) of the lectures and lecture resources can affect the team process and output. When the team members do not understand the lectures or lecture resources, it negatively impacts the team process. Students will likely spend some time understanding task-related lectures and lecture resources before working on the tasks. Additionally, in the bid to provide as much information as required for learning, there were some complaints of the lectures having too much content, which can be confusing and affect the students' learning process. To improve the lecture intervention when following this hackathon approach, we suggest more consideration of the balance between the quantity of theory provided and its applicability in the hackathon tasks and provide ample opportunity for feedback sessions where educators can further discuss the lectures and how it relates to the hackathon tasks.
Secondly, we expect that familiarity between members of teams formed during online education and hackathon events will likely have low team familiarity levels. Thus hackathon interventions must account for this and work towards improving collaboration regardless of familiarity levels. Also, introducing more interesting real-life use-cases and possible hands-on collaboration with the industry in the use-cases provided can help to boost student and team interest and participation.

\subsection{Limitations}
There are certain limitations associated with this particular study design. We developed specific interventions and studied teams participating in the hackathon events as part of a particular cybersecurity course. First, it is not possible to generalize our findings beyond the context of our specific course since another study on a different course and with different use-cases might yield different results. However, the point of our work is not generalization but rather to evaluate the interventions, report findings and provide suggestions on how to handle hackathon integration to online-led courses.
Second, the sample size of our study may include bias based on our selection for analysis. However, we selected a cross-section of student teams covering high, middle and low grade levels and meeting other requirements such as team size and questionnaire response completeness.
Third, two out of three researchers conducting the study were involved in the hackathon planning, execution, and course grading, which can introduce bias to the reported findings by the students. Additionally, the post-hackathon questionnaire for all three hackathon events was not anonymous, also introducing bias to the reported findings.
However, one researcher had no involvement in executing the hackathon and refrained from interfering during the hackathon and the course until analysis of the collected data began. We began our analysis of collected data after all three hackathon events and the cybersecurity course was completed to prevent bias in grading the students involved in our analysis based on how they reacted to our intervention style. Since we compare within the students, such bias will affect all students equally. Furthermore, we abstained from making causal claims in our analysis; instead, we provide a rich description of students and teams' observed behaviour and reported perceptions.
Lastly, there might be a bias in reporting and analyzing the open-ended questions; however, we did not generalize or draw final conclusions but used the responses as potential explanations to our findings.

\section{Concluding Remarks}
This paper reported findings from an action research study of six (6) teams at a series of educational hackathons integrated into an online cybersecurity course. The study aimed to show how educators can support teamwork and collaboration, maintain student participation and interest, and encourage learning-by-doing throughout the course through educational hackathons in an online context. We introduced the \textit{lecture}, \textit{feedback} and \textit{team management plan} interventions to achieve our hackathon goals.

Our findings indicate that these interventions helped the students to achieve the course learning outcomes by knowledge sharing through \textit{lectures}, guidance on applying knowledge gained primarily through \textit{feedback}, improving efficiency in completing the given hackathon tasks through the \textit{team management plan}. Collectively, these interventions improved the team collaborative power and maintained interest and participation in the online course, thereby addressing challenges faced with online instruction. Our results also point to suggestions useful for future iterations of the hackathon format.


\bibliographystyle{ACM-Reference-Format}
\bibliography{ref}
%
%

%
\newpage
\appendix
\section{Appendix}\label{sec:app}
\begin{table}[ht]
\caption{Post-Hackathon Questionnaire Instrument}
    \label{tab:Qinstruments} 
    \centering
\begin{tabular}{|p{\textwidth}|}
\hline
\cellcolor{lightgray}
Team familiarity (based on Filippova\textit{ et al.}\cite{filippova2017diversity}), anchored between not at all and completely. \\ \hline
I know my team members well. \\
I have collaborated with some of my team members before. \\
I have socialized/met with some of my team members outside of work/school before. \\ \hline
 \cellcolor{lightgray}
Team process (based on Bhattacherjee\cite{bhattacherjee2001understanding}), anchored between 1 and 5. \\ \hline
I am satisfied with the work completed in my project.\\
I am satisfied with the quality of my team's output.\\
My ideal outcome coming into my team was achieved. \\
My expectations towards my team were met. \\ \hline
 \cellcolor{lightgray}
Perceived satisfaction with team process (based on Filippova\textit{ et al.}\cite{filippova2017diversity}), anchored between strongly disagree and strongly agree. \\ \hline
(1) Inefficient to (5) Efficient \\
(1) Uncoordinated to (5) Coordinated \\
(1) Unfair to (5) Fair \\ 
(1) Confusing to (5) Easy to understand \\ \hline
 \cellcolor{lightgray}
Team goal clarity (based on Nolte\textit{ et al.,}\cite{nolte2018you}) anchored between strongly disagree and strongly agree.
\\ \hline
I was uncertain of my duties and responsibilities in this team.\\
I was unclear about the goals and objectives for my work.\\
I was unsure how my work relates to the overall objectives of my team.\\ \hline
 \cellcolor{lightgray}
Perception of team participation and voice (based on Nolte\textit{ et al.}\cite{nolte2018you}) anchored between strongly disagree and strongly agree.
\\ \hline
 Everyone had a chance to express her/his opinion.\\ 
 The team members responded to the comments made by others.\\
 The team members participated very actively during our collaboration. \\ 
 Overall, the participation of each team member was effective. \\ \hline
 \cellcolor{lightgray}

Perception of the usefulness of the interventions (based on Sauro\cite{sauro2011measuringu}) anchored between strongly disagree and strongly agree.
\\ \hline
 Using the \textit{[intervention]} enabled me to accomplish tasks more quickly. \\
 Using the \textit{[intervention]} improved my team's performance. \\
 Using the \textit{[intervention]} increased my productivity in the hackathon. \\
 Using the \textit{[intervention]} enhanced my effectiveness in my team. \\
 Using the \textit{[intervention]} made it easier to complete my \textit{[hackathon]} solution. \\
 I found the \textit{[intervention]} useful in my team. \\ \hline
 \cellcolor{lightgray}
Learning outcome measured students' perception of achieving the course's learning outcomes, perceived learning process, and learning through problem-solving; anchored between strongly disagree and strongly agree. \\ \hline
%
The hackathon events allowed me the opportunity to design secure systems and software.\\
The hackathon activities made my learning experience more productive. \\
The lectures given were geared to promote my understanding. \\ 
There were enough opportunities during the course to find out if I clearly understood the course material. \\
The \textit{[interventions]} given were appropriate and geared to promote my understanding. \\
\hline
\end{tabular}
 \vspace{-10pt}
\end{table}

\end{document}